# Designing Cybersecurity Awareness Solutions for the Young People in Rural Developing Countries: The Need for Diversity and Inclusion

**Farzana Quayyum and Giske Naper Freberg**
*Norwegian University of Science and technology (NTNU)*
*farzana.quayyum@ntnu.no; giskefreberg@hotmail.com*

**Abstract**
*Cybersecurity challenges and the need for awareness are well-recognized in developed countries, but this still needs attention in less-developed countries. With the expansion of technology, security concerns are also becoming more prevalent worldwide. This paper presents a design and creation research study exploring which factors we should consider when designing cybersecurity awareness solutions for young people in developing countries. We have developed prototypes of mini-cybersecurity awareness applications and conducted a pilot study with eight participants (aged 16-30) from Gambia, Eritrea, and Syria. Our findings show that factors like the influence of culture and social constructs, literacy, and language competence, the way of introducing cybersecurity terms and concepts, and the need for reflection are essential to consider when designing and developing cybersecurity awareness solutions for target users in developing countries. The findings of this study will guide future researchers to design more inclusive cybersecurity awareness solutions for users in developing countries.*

**Keywords**: Cybersecurity awareness · Developing countries · Young people · Human factors of cybersecurity · Diversity and inclusion.

## 1.0 Introduction

The use of information and communications technologies is crucial for the advancement of economies in developing countries (Romanoff et al., 2018; Hatakka et al., 2020). The recent COVID-19 pandemic has accelerated the use of technologies worldwide. During the pandemic, the need for and importance of technological development in developing countries have again come to light, underlining the digital gaps and how much we need to bridge this gap. Although providing internet connectivity and digital technologies is a crucial part of the development, making the technologies available only is not enough (Aruleba & Jere, 2022). Training, adapting, and learning the correct usage of the technologies are just as necessary as availability.

In cybersecurity awareness research, many research studies have been conducted to design and develop educational resources for children and adults (e.g., Giannakas et al., 2015; Zhao et al., 2019; Bioglio et al., 2018). However, most of these studies are done in Western and developed countries, keeping the user groups from these specific parts of the world in mind. Developing cybersecurity awareness solutions for users in developing countries and least developed countries (LDCs), still lacks attention. The importance of online security issues grows along with the development of ICTs and the rise in online interactions. Limited studies have explored cybersecurity issues in developing countries (e.g., Owusu et al., 2019; Da Veiga et al., 2022), but the amount of research is inadequate and lacking. The users' needs and facilitating learning factors can vary depending on the socio-economic diversity of the user groups, which are very different in developing countries compared to Western and developed nations.

Access to technologies, the internet, and digital resources are correlated with the economic status of an individual and a family (Bernadas & Soriano, 2018). Illiteracy, poverty, and a weak economy are common phenomenons in LDCs; thus, people's familiarity and competence with the digital world are also weak compared to those living in developed countries. However, the accessibility of the internet and digital devices is increasing globally, including the developing countries. So, with this increased engagement with digital technologies, the need for cybersecurity awareness also emerged. However, considering that people in developing countries have limited knowledge about technologies and maybe even a limited level of education, designing and developing cybersecurity educational resources for them is a challenge. Thus, we address the following research question in this study: *What factors should we consider when developing cybersecurity awareness solutions for users in developing countries?*

The outline of this paper is as follows: We discuss the background of this study and related works in Section 2, whereas in Section 3, we present our research methodologies. In Section 4, we present the results of this study, followed by a discussion of the findings in Section 5. In the end, we conclude our paper in Section 6, highlighting the future directions of our work.

## 2.0    Background and Related Work

"Developing countries" is a collective term that refers to countries that suffer from a weak economy and various socio-economic difficulties that constrain the population's prosperity in terms of development, living standard, and welfare. Though developing countries can be further divided into subgroups such as LDCs, low-income countries, high-income developing countries, and so on, for this study, we mainly consider and talk about young people (aged up to 30 years) from the rural areas of the LDCs. However, in this study, we use both the terms developing countries and LDCs simultaneously (as LDCs are a subgroup of developing countries).

The widespread accessibility and decreasing cost of technologies created a significant increase in interest in using ICTs in the context of developing countries (Walsham, 2017). According to Poushter et al. (2018), the smartphone ownership gap between developed and developing nations is narrowing. Nevertheless, even though there is rapid growth in the numbers of internet access and digital media usage, developing countries are still facing an immense challenge in providing digital literacy and security training for their populations. Various research studies have shown that measures should be taken to escalate the cyber awareness level among people in developing countries (e.g., Chang & Coppel, 2020; Adelola et al., 2015). Yet, cybersecurity awareness in developing countries is still an under-researched area and faces various challenges, including the lack of comprehensive initiatives, limited governmental support, inadequate cybersecurity curricula or extramural activities, limited budgets, and lack of resources (Kortjan & Von Solms, 2013; Von Solms & Von Solms, 2015).

Education and training on cybersecurity are not new and have been a focus in Western and well-developed countries for many years. Several tools, games, courses, training materials, and other learning activities are available to help people learn about cybersecurity and be aware of the issues internet usage can bring. Many of these available resources are targeted toward children and teenagers as they are one of the most vulnerable groups to targeted cyber-attacks. Example of such educational resources includes games like Be Internet Awesome by Google[1], CyberCIEGE (Cone et al., 2007), interactive books like Cyberheroes (Zhang-Kennedy et al., 2017) and so on. There are various educational solutions for adults as well, including (Raisi et al., 2021; Bacud & Mäses, 2021). However, most of these studies are done, and resources are developed with an audience in mind, which is located in developed countries. Many children and young people born or living in developed countries are growing up in a digital world surrounded by many advanced technologies and devices. Young people in developing countries, on the other hand, may not have had the same environment and exposure to the digital world. Many get access to the internet and own personal devices like mobile phones, smartphones, and computers when they become young adults or even later. Thus, they might embrace the digital world for the first time without any prior knowledge. Moreover, cybersecurity issues may not be exactly the same in all countries (Walsham, 2017).

Maoneke et al. (2018) conducted a study to identify cyberspace risks for adolescents in Namibia. The researcher found many risks, including exposure to inappropriate content (such as pornographic content, harmful content, and violent videos), cyberbullying, stranger dangers, privacy violations, etc., as the common risks the young generation faces online. Some other studies have also explored the cybersecurity risks and the need for awareness in the context of developing countries (e.g., Kritzinger et al., 2018; Naidoo et al., 2013). Bernadas and Soriano (2018) investigate the diversity of internet connectivity in developing countries and the link of privacy behaviors of the youth with the diversity of connectivity in terms of information literacy. Von Solms and Von Solms (2015) propose a cyber-safety curriculum for teachers in junior or primary schools using open educational resources.

---

[1] https://beinternetawesome.withgoogle.com/en us

Nevertheless, many of the existing studies conducted in the African region contribute to the cybersecurity awareness needs from a strategic and policy-making perspective. As the majority of the resources are created and produced considering the users of developed nations, there is a shortage of research into the habits and user needs when providing cybersecurity educational solutions for users in developing countries and rural areas.

## 3.0    Methodology

For this study, we have used a design and creation research strategy focusing on developing educational solutions for cybersecurity. We have developed apps to help the target audience learn about three cybersecurity topics: online contacts, password security, and privacy. This paper presents the results of a pilot study where we test our prototypes of mini-educational applications (apps). To develop the prototypes for this study, we have collaborated with a Norwegian company named Leap Learning[2] that develops educational solutions for children and adults with an aim to bridge the educational and digital divides around the world. This company is actively working with many developing countries in Sub-Saharan Africa. As they already have a connection and functional platform in that region, we collaborated with Leap Learning to construct initial prototypes of the educational apps and to conduct a pilot testing with participants from a Sub-Saharan African nation (i.e., the Gambia). Here, we would like to clarify that our main focus for this study is to identify the needs of our targeted user groups and understand the issues we must consider when designing cybersecurity educational solutions for our specific user domain. Therefore, we focused only on exploring the user needs using the existing design templates and development platform of Leap Learning. In the following section, we describe our methodology and the prototypes.

### 3.1 Development Platform

As mentioned earlier, the prototypes for this study were created using L's design templates and developmental platform. Thus, our prototypes contain designs and functionalities pre-existing in the development platform of Leap Learning. After implementing the templates in the development platform, we added content and relevant illustrations to each app. After the initial implementation, changes could be made to each app, such as updating the number of activities, levels, features, locations and adding new material.

### 3.2 Content Elicitation

Based on the findings from relevant studies, we identified the common cybersecurity risks that young people face nowadays, such as privacy, password security, online strangers, cyberbullying, phishing, identity theft, and many others. Quayyum et al. (2021) conducted a literature review presenting a list of cybersecurity risks relevant to children, adolescents, and young people. Svabensky et al. (2020) also conducted a literature review to investigate what cybersecurity topics are frequently explored and which are underrepresented in academic papers. The study by Svabensky et al. also explored the common cybersecurity teaching practices in the context of the USA. For our pilot study, we chose not to focus on all cybersecurity issues in one iteration cycle in accordance with the design and development research strategy's iterative cycle principles. We have selected three of the risks (namely, online contact, privacy, and password security) identified in the literature for the initial iteration and this pilot study because users of different ages worldwide often encounter them. To formulate the questions for the apps, we mainly took inspiration from the cybersecurity awareness curriculum by Google[3]. In addition, findings from the literature (Quayyum et al., 2021; Svabensky et al., 2020; Zhang-Kennedy et al., 2021) were used as inspiration. Since our target audience was from rural areas in developing countries and might not have great digital competence, we kept the prototypes basic and informative. We decided to make the apps available in English to reach a wider audience. In the Gambia, where English is the official language, the majority speak, write, and understand English orally to a certain level.

---

[2] https://leaplearning.no/

### 3.3 Functionality Elicitation

Our selected design template uses different approaches and game elements, including quizzes, storytelling (to provide information), and prioritizing. From the literature (e.g., Zhang-Kennedy et al., 2017; Giannakas et al., 2015; Baciu-Ureche et al., 2019), we have seen that storytelling, quizzes, visualization, and the inclusion of gamification elements were among the common functionalities and practices applied to raise cybersecurity awareness. Utilizing the same approach for all the apps could bore the participants and make them feel less driven to test out the apps, so we have employed these three functionalities for the prototype applications. Another reason behind different approaches is that they may help us determine which approach works best for the intended audience.

### 3.4 Prototype Development

For all three topics (online contact, privacy, and password security), we have developed one informative app and one quiz app each. For the topic of password security, we have developed one additional app using the priority and sorting approach (to test the user's understanding of password strengths). Thus, we have developed and tested a total of seven mini-apps (using three approaches). The approaches are further described below.

**The informative app.** An informative app informs users about a topic by giving useful information in a textual format. This app format includes short sentences accompanied by an illustration and aims to explain a relevant aspect concisely and descriptively. Each sentence on the app tells users about security issues that may arise while using the internet and the possible consequences. Figure 1 displays two screenshots from the working prototype of the informative app on online strangers.

---
[3] https://beinternetawesome.withgoogle.com/en us

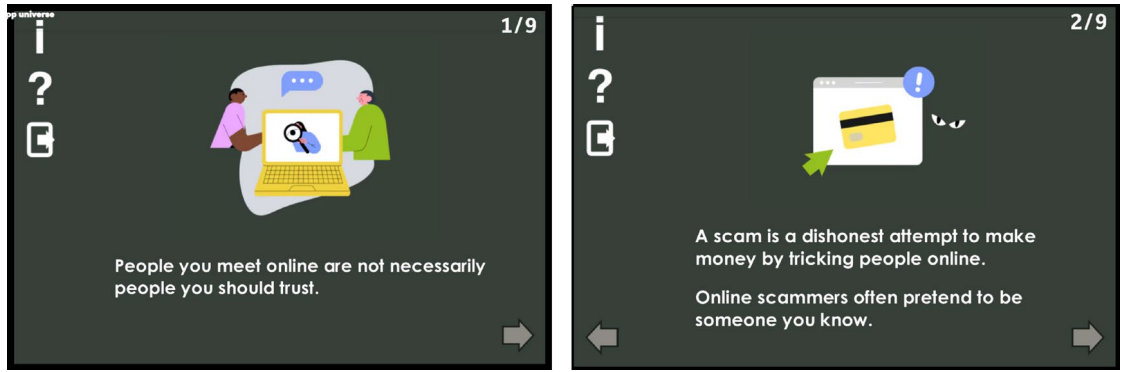

Figure 1. Informative app, focused on online strangers.

**The quiz app.** To challenge the user and test if they have understood and learned from the informative app, we included a quiz inspired "Select Sentence" app. This app asks the user a question, and the user is given three alternative answers to choose from. This app features a drag-and-drop functionality where the user needs to drag the correct answer into the answer slot. When the user drags the correct answer to the answer slot, it will turn green and play a celebratory sound. Figure 2 displays an example of the question with the view of the interface before and after answering.

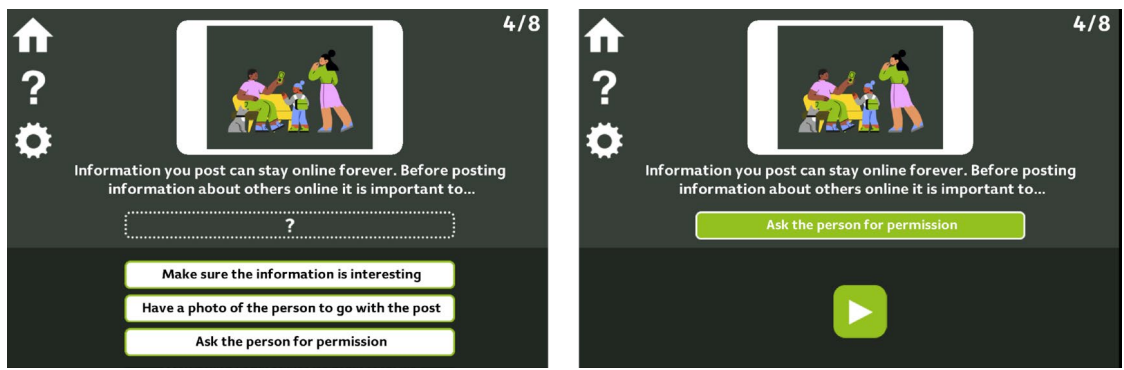

Figure 2. Quiz app, focused on sharing personal information online.

**Priority app.** The next approach is priority apps, where the user prioritizes the given alternatives from "best to worst" or "worst to best." We have used this approach to make an app focused on password security; we asked the users, for example, to arrange some given passwords according to their strengths in terms of security. This app also incorporates drag-and-drop functionality. Figure 3 displays an implemented view of a priority task before and after the correct prioritization.

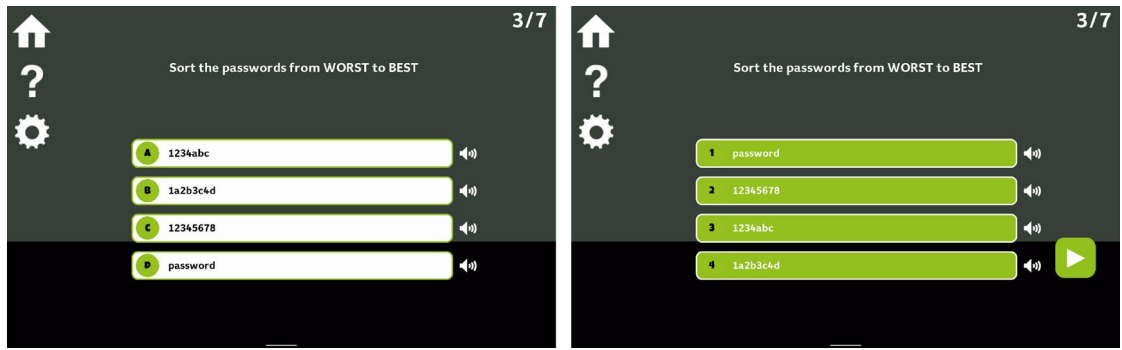

**Figure 3. Priority app, focused on password security.**

### 3.5 Participant Recruitment

We recruited six participants (P1-P6) from the Gambia for the pilot study. In addition, we recruited two participants (P7 from Syria and P8 from Eritrea) living in Norway (who had recently arrived in Norway as refugees from their respective countries). To recruit the participants in Norway, we contacted an education center for adolescents and young adults, which offers primary and high school-level teaching for immigrants and refugees. Gambia and Eritrea are both countries listed on the UN's list of LDCs, while Syria is listed as a developing country. The participants in the Gambia were primarily students. The representatives from Leap Learning contacted the students through their schools and recruited them. An overview of the participants can be seen in Table 1. The gender distribution of the participants was five males and three females. The age range of the participants varied from 16 to 30 years. To comply with the ethical regulations, we took approval from the Norwegian Agency for Shared Services in Education and Research (SIKT, formally known as NSD)[4] to collect personal data from the users. Before the user testing, we explained the purpose and objectives of the study to each participant. All the participants were required to sign an informed consent. The participants were also informed about their right to withdraw from the process.

### 3.6 Data Collection

During the pilot study with the prototypes, we observed the participants interacting with the apps, followed by an interview. The observation was used to discover whether the apps were simple for the participants to use and whether they had any issues while attempting them independently. Another goal of this testing and observation was to understand the users' needs for future development. Due to time limitations, we tested each of the apps with five participants among the eight. Our strategy was for each participant to test the apps on two cybersecurity topics. Afterward, we conducted semi-structured interviews so the participants or the interviewer could ask follow-up questions if needed. The purpose of the semi-structured interviews was to know the demographic information about the participant and the participant's feedback and reflections on the tested applications. Demographic data included age, country of origin, and the user's previous experience with digital media. While obtaining the participants' feedback after interacting with the working prototype, we asked them if they had any difficulties understanding how to use any of the apps, if they learned anything new when using the apps, and what they enjoyed about the apps or not.

---

[4] https://sikt.no/en/about-sikt

### 3.7 Data Analysis
The collected data consisted of notes and comments from the observation phase and recorded audio files from the interviews. We performed a qualitative analysis with the interview transcripts and observation data. For the data from the observations, we conducted a qualitative categorization analysis following Preece et al. (2015). This approach involves looking for incidents and patterns in the data. When analyzing the transcripts and observation notes, we looked for similar facts or themes that appeared multiple times and would be helpful for us to improve the apps in the future. The themes and categories mainly evolved around participants' reflections and perceptions of the apps' usability. We further present the identified themes in Section 4.

## 4.0   Results
To give an overview of the participants' digital competence, we start by summarizing the results of participants' digital usage and whether or not they had access to digital devices and the internet in their daily lives. Most participants owned a smartphone and were familiar with social media, as shown in Table 1.

| ID | Age | Origin | Digital device | Internet access | Social media | Have Email |
|---|---|---|---|---|---|---|
| P1 | 16 | Gambia | No | Only at school | No | No |
| P2 | 18 | Gambia | Smartphone | Yes | Facebook, WhatsApp, Twitter | No |
| P3 | 17 | Gambia | Phone, computer | Yes | Instagram, WhatsApp, Twitter, Google | Yes |
| P4 | 23 | Gambia | Phone | Yes | No | Yes |
| P5 | 23 | Gambia | Phone, computer | Yes | Facebook, WhatsApp, Instagram | No |
| P6 | 30 | Gambia | Phone, computer | Yes | Instagram, Facebook, WhatsApp, Snapchat | Yes |
| P7 | 23 | Syria | Smartphone, computer | Yes | WhatsApp, Facebook | Yes |
| P8 | 28 | Eritrea | Smartphone, | Yes | Facebook, WhatsApp, Google | Yes |

**Table 1. Overview of the participants.**

## 4.1 Participant's Interactions with the Apps

The key findings on how the participants interacted with the apps' features during the observation sessions are presented in this section. We outline our observations in accordance with the various app functioning categories.

**The informative apps.** Overall, the participants interacted with the apps on the tablet seemingly well; they did not struggle with the app's navigation and could proceed to the next screen without difficulty. However, the main issue observed in the informative apps was understanding certain words and terms.

**The quiz apps.** Most participants who tested the quiz apps understood the "drag-and-drop" concept. However, participants P2, P4, and P7 needed some time to understand the "drag-and-drop" functionality and tried to push the answers instead of dragging them to the marked "answer" box. On the other hand, participant P7 seemed to find the drag-and-drop exciting and expressed a joyful "yes!" expression when answering a question correctly. All the participants managed to navigate the following tasks without any noticeable issues.

**The priority app.** The priority functionality was only implemented for the password security topic. We observed that all five participants struggled to understand how the functionality worked for this app. Participants tried to press the buttons instead of dragging them into the correct order. But all participants seemed to get it after some trial and error. However, we occasionally placed the order of alternatives in the right order from the beginning of the game. In these cases, the alternatives had to be moved back and forth to turn green and registered as correct. This functionality confused the participants.

**Other issues**. In addition to the above-mentioned observations related to each functionality type, we also noticed some other issues in general when the participants were trying the apps. It was observed that five out of eight participants found it difficult to exit the app. Participants P1, P2, P3, P4, and P8 did not understand the exit symbol on the button. Eventually, the participants understood the exit functionality after receiving guidance from the observer. An- other issue was related to the content and language. Some participants did not seem to struggle with understanding the content, while others struggled with certain words and terms (depending on their English language competence). There were many online security-related concepts or related words that some participants struggled to pronounce and understand the meaning or the idea. Malicious, extortion, digital footprint, and catfishing are some examples of such words the participants found difficult. The term "friend request" was also new to the participant who had no social media.

## 4.2 Participants' Reflections on the Apps

Participants' reflections on the apps were diverse. Some participants thoroughly reflected on their experiences, while some were comparatively reserved in their responses. When asked if they had previously received any cybersecurity training or education, all eight participants answered "no". As a result, their interactions with the apps were their first encounter with cybersecurity topics in an educational setting. Regarding learning from the applications, every participant gave positive feedback. For example, participants P1, P2, P5, and P8 acknowledged how the apps taught them about privacy and the risks of sharing personal information with strangers. Participants P2, P3, P6, and P7 said that the applications assisted in learning secure password practices, according to P7 *"I learned that the passwords should be more difficult than just numbers. So here is 1234, or maybe the date when you were born. I used that before. Especially that 12345678. That one is the easiest and everybody can, for example, if they had the phone (my phone) in their hands they could try with that number 123456. So, it's really easy. We should never use it."*. Participant P5 also mentioned scamming and how online strangers can scam people if they get private information about the individual. Other participants also agreed about learning from the apps but did not elaborate on what and how they learned.

Along with asking about what they learned from the apps, we also asked the participants if they were familiar with any information presented in the apps (i.e., prior to testing the apps). Even though the participants had not received formal training, some had a basic understanding of online security issues. For instance, participants P2, P3, and P7 mentioned they knew they should not share passwords with others; participants P1 and P8 knew they should not share personal information like phone numbers or photos with strangers. Participant P6 also added about having knowledge about scamming risks as she experienced it in real life.

In addition to learning experiences, we further asked the participants which topic or app they enjoyed most. All the participants who tested the password app mentioned that password security was their favorite topic. The priority app on passwords (arranging the given passwords) seemed to be a favorite of multiple participants. Participant P6 said, *"When it came to finding the rating from the least to the most important, I found that really interesting because it was testing me, and it was a little difficult"*. Participant P7 expressed that *"It is more fun and has more information about learning how to use the passwords"*. We note that we have not rephrased the quotes presented in this paper; instead, we are presenting them as the participants stated (while fixing any grammatical errors).

## 4.3 Participants' Perception of Usability

When asked if they had any difficulties using the different apps, the first five participants (P1-P5) reported that they did not have any challenges and perceived the apps as simple. Participant P6 reported having a few difficulties but provided no further explanation about what kind of difficulties. Another participant, P8, mentioned some struggles figuring out how to enter and exit the application. Participant P7 provided more detailed feedback regarding the priority app on password security. The participant found it strange that once he had chosen one alternative and everything turned green, he could not go back and change his answer. He also explained that he would have liked to have more feedback when getting the answers correct, "like an OK or something" as he stated.

### 4.4 Participants' Perception of the Apps' Language

When asked how they understood the language in the apps, the vast majority of participants replied, *"there were a few words I did not know"*. A spelling error was discovered in one of the apps by participant P6, which shows that the language was easy enough for this participant to identify any mistakes. Nevertheless, it also depended on the proficiency level of the participant himself. The general response was that the sentences were "not too long and difficult." Since participant P4 was not proficient in reading English but was proficient in speaking and understanding the language, the participant had trouble reading the content from the apps. The participant, however, remarked that the language was "really easy" to comprehend after being assisted by the observer to try the apps and read aloud the app's contents.

## 5.0 Discussion

In this section, we discuss the findings of this study and the lessons we learned about what factors to consider when developing cybersecurity awareness solutions for users in developing countries, followed by the potential limitations.

### 5.1 Existing Cybersecurity Knowledge and Experience

In Section 4.2, we mentioned that some participants mentioned having an aware- ness of privacy, passwords, and scams, even though they said they did not get any formal education or training on cybersecurity before participating in our study. From the interviews and conversations with the participants, it is apparent that they became aware of these security issues from their real-life experiences. For example, three participants (P2, P5, and P6) mentioned experiencing scamming attempts on social media. The participants who have smartphones or computers reported knowing about passwords, which they seem to have achieved due to using these devices. However, even though the participants know they should not share their passwords, they still lack awareness about secure passwords, recommended password-strength practices, etc. Overall, the results of this study suggest that young people in rural developing nations are capable of comprehending cybersecurity challenges if given the right guidance and education. Even though some of the participants had trouble understanding certain terminology and words used in the apps, they were able to understand the meaning after being explained.

## 5.2 Factors We Should Consider

From the results, we can say that despite living in rural areas in developing countries with limited digital access and opportunities, young people are exposed to online security threats, as highlighted by earlier studies as well (e.g., Maoneke et al., 2018; Von Solms & Von Solms, 2015). Thus, emphasis on cybersecurity training is needed to make them responsible and sensible users of the internet and digital tools and technologies. Nevertheless, depending on the available access to technologies and differences in the learning environment and culture of the users in different parts of the world, our proposed solutions should be relevant and inclusive for our targeted audience. From observing the participants and talking to them, we have identified some aspects that need to be considered when developing digital cybersecurity education platforms for an audience in underdeveloped areas of the world. In the following section, we present these identified factors.

**Influence of culture and social constructs.** It is crucial to adjust cultural and local content when communicating new information, as we discovered from the pilot study in the Gambia. To reach a wider audience, we created our apps in English and used language that could be easier for the general public to understand. However, several terms were used in the apps to present the cybersecurity topics, which the participants struggled to understand or could not relate to. One participant, for instance, was confused about what a "pet" was (the term was used in the password app). Others could not have understood the phrases used to describe banking and financial services (to describe financial fraud or scam) because having a bank account is uncommon in rural areas. Preece et al. (2015) emphasized the importance of formulations, language, or jargon when formulating the interview guide. This aspect became evident in our study when the participants were asked the following question: "What gender do you identify as?". In developed countries, especially in Western culture, asking a gender question may seem correct or normal. However, cultural customs may be far more traditional in rural areas of least-developed nations like the Gambia; gender roles and identification in these nations have not been questioned. Therefore, the participants were quite unfamiliar with a question that had been phrased in this way. To ensure that each participant understood the question, we had to repeat it and rephrase it.

Social constructs and phenomena in rural areas differ from those in Western countries. Particularly in rural areas, society is more traditional than in urban areas. Thus, topics that deal with, for example, sexuality and gender identification are rarely or never a topic of conversation. Having a pet in a household is not a common phenomenon. These terms refer to phenomena and constructions common in societies in developed countries. So, culture and social context are essential to consider; each country is different regarding cybersecurity awareness, the extent of technological advancement, cybercrimes, and their ability to handle these matters at a societal or national level. Chang and Cobbel (2020) also highlighted in their study how a country's context and social structure could create challenges and influence the effectiveness of a cybersecurity awareness program.

We have mentioned earlier that in some cases, it was evident from the observation that some participants found it challenging to use some of the apps' features or answer questions. However, when asked about their struggles during the interview, they did not reflect on these difficulties and expressed that there were no challenges in understanding the apps and that the process was "easy." This behavior might indicate that the participants were concerned with answering the questions in a manner they believed would be pleasing or expected by the interviewer rather than being straightforward with their opinions.

**Literacy and language competence**. The language and literacy levels of people in the least developed countries are not necessarily correlated to their age. It depends on how much schooling they have received throughout their lives. Though English is the official language of the Gambia, and people may have a certain level of proficiency in the English language, they still have issues com- prehending the concepts in English. During our study, some participants needed help in reading and understanding the information provided in the apps because some participants could understand oral English but were not good at reading and understanding. To address this gap between oral understating and literacy levels, including sounds and audio of written contents could be part of the solution. The audio could accompany the textual visualization, giving illiterate people equal access as illiteracy is a common problem in underdeveloped countries and areas. Another interesting example was observed while interviewing participant P1. When asked about the learning outcomes from the apps, participant P1 replied by reciting some text from the app word by word. However, she did not reflect upon what she had read and the context of the information. This can indicate that she was more concerned with remembering the words and sentences by heart rather than comprehending what she was reading.

**Introducing cybersecurity terms and concepts**. Cybersecurity is a field where there are many special terms and terminology used. Introducing these special terms and terminologies to the target audience is challenging if the literacy level is low or absent. As stated in Section 4.1, "scam," "malicious," "catfishing," "blackmail," "extortion," and "digital footprint" are examples of cybersecurity terms the participants struggled to interpret. Nevertheless, these are some common cybersecurity risks and are important to include when educating about cybersecurity. Identifying ways to help people understand these cybersecurity topics is essential, even if they have no or low literacy level. One suggested approach could be to connect with real-life situations and examples to explain the terms and help them understand the consequences and the preventative measures. While interviewing participant P2 after testing the apps, she shared her experience about an episode when she almost got scammed by a stranger on Facebook. She spontaneously recounted this occurrence with the interviewer because she could see the similarities between this incident and the information provided in one of the apps. This way, it was easier for this participant to under- stand the concept of a scam, how it works, and the consequences. Thus, relating the cybersecurity concepts with real-life scenarios will help increase awareness even if the target audience lacks formal education. Also, a gradual introduction of the terms will allow the user to process the information gradually by dividing the topics and relevant information into levels based on the complexity of the topic and related difficulties. When a user acquires a foundational understanding of cybersecurity topics, introducing difficult words can follow at a higher level.

**Reflection on the actions**. One of the main takeaways from the study by Adelola et al. (2015) was the importance of measuring and testing the users' knowledge and ability level. The purpose of providing comprehension tests is to encourage users to reflect and think critically about the data presented. In addition, it can give an indication to the educator whether or not the information fits the needs of the target audience. Our quiz-inspired prototype included a "Select Sentence" functionality and a prioritization functionality for the password security topic. These functionalities were implemented using a drag-and-drop feature. Both of these functionality types aimed to test whether or not the user comprehended the information provided in the informative apps. The participant had to finish the task by giving the right response in order to go on to the next one. The apps gave users feedback on their answers by becoming green when the right response or solution was supplied, as described in Section 3.4. However, it was found throughout the observations that this feature pushed some users to attempt several solutions until they found the right one rather than reflecting before choosing an answer. Giving the user one or two chances to submit the right response before moving on to the next task may help us avoid this problem in future iterations. The user can receive feedback or a score on the tasks they completed correctly and incorrectly once all the tasks have been completed. This strategy might also be more motivating and promote focus and reflection.

**5.3 Limitations and future work**

The primary limitation of this study is that our sample size was small. We acknowledge that studying a larger sample size would have given more versatile results and related findings. As we have mentioned earlier, this is a pilot study; thus, we plan to make further improvements to the existing prototypes and continue with more studies in multiple phases. Another limitation is that the apps are primarily written in English. The majority of young people in The Gambia can understand spoken English. Still, as we found during the study, they may need to develop their written English skills to fully comprehend the context and significance of the app's content. As stated earlier, we chose to use English to reach a wider audience and make our study applicable to people from different countries.

# 6.0    Conclusion

This paper presents the findings of our study exploring how we can create inclusive and efficient cybersecurity awareness solutions for young users in rural developing countries. In Western and developing or less developed nations, the "cyber maturity" level varies. To better meet needs of diverse user throughout the world, we need to be more inclusive in our designs and overall cybersecurity awareness research. The positive findings of this study suggest that, when given the right tools and learning opportunities, young people from developing countries or LDCs are well capable of understanding and comprehending the concepts and etiquette of cybersecurity, even though they may not currently have access to institutional higher education and lack experience using digital tools. This pilot study provided us with a number of suggestions for alterations and enhancements to the current prototypes. As part of our ongoing research and future work, we aim to improve the current apps and continue more studies to evaluate the prototypes further and explore the issues.

## Acknowledgements

We thank Professor Letizia Jaccheri for all her support and valuable guidance throughout the work.